\documentstyle[12pt]{article}

\def\hybrid{\topmargin -20pt	\oddsidemargin 0pt
	\headheight 0pt	\headsep 0pt
	\textwidth 6.25in	
	\textheight 9.5in	
	\marginparwidth .875in
	\parskip 5pt plus 1pt	\jot = 1.5ex}

\hybrid

\catcode`\@=11

\def\marginnote#1{}
\newcount\hour
\newcount\minute
\newtoks\amorpm
\hour=\time\divide\hour by60
\minute=\time{\multiply\hour by60 \global\advance\minute by-\hour}
\edef\standardtime{{\ifnum\hour<12 \global\amorpm={am}%
	\else\global\amorpm={pm}\advance\hour by-12 \fi
	\ifnum\hour=0 \hour=12 \fi
	\number\hour:\ifnum\minute<10 0\fi\number\minute\the\amorpm}}
\edef\militarytime{\number\hour:\ifnum\minute<10 0\fi\number\minute}

\def\draftlabel#1{{\@bsphack\if@filesw {\let\thepage\relax
   \xdef\@gtempa{\write\@auxout{\string
      \newlabel{#1}{{\@currentlabel}{\thepage}}}}}\@gtempa
   \if@nobreak \ifvmode\nobreak\fi\fi\fi\@esphack}
	\gdef\@eqnlabel{#1}}
\def\@eqnlabel{}
\def\@vacuum{}
\def\draftmarginnote#1{\marginpar{\raggedright\scriptsize\tt#1}}

\def\draft{\oddsidemargin -.5truein
	\def\@oddfoot{\sl preliminary draft \hfil
	\rm\thepage\hfil\sl\today\quad\militarytime}
	\let\@evenfoot\@oddfoot	\overfullrule 3pt
	\let\label=\draftlabel
	\let\marginnote=\draftmarginnote
   \def\@eqnnum{(\theequation)\rlap{\kern\marginparsep\tt\@eqnlabel}%
\global\let\@eqnlabel\@vacuum}  }

\def\preprint{\twocolumn\sloppy\flushbottom\parindent 2em
	\leftmargini 2em\leftmarginv .5em\leftmarginvi .5em
	\oddsidemargin -.5in	\evensidemargin -.5in
	\columnsep .4in	\footheight 0pt
	\textwidth 10.in	\topmargin  -.4in
	\headheight 12pt \topskip .4in
	\textheight 6.9in \footskip 0pt
	\def\@oddhead{\thepage\hfil\addtocounter{page}{1}\thepage}
	\let\@evenhead\@oddhead	\def\@oddfoot{}	\def\@evenfoot{} }

\def\numberbysection{\@addtoreset{equation}{section}
	\def\theequation{\thesection.\arabic{equation}}}

\def\underline#1{\relax\ifmmode\@@underline#1\else
	$\@@underline{\hbox{#1}}$\relax\fi}

\def\titlepage{\@restonecolfalse\if@twocolumn\@restonecoltrue\onecolumn
     \else \newpage \fi \thispagestyle{empty}\c@page\z@
	\def\thefootnote{\fnsymbol{footnote}} }

\def\endtitlepage{\if@restonecol\twocolumn \else \newpage \fi
	\def\thefootnote{\arabic{footnote}}
	\setcounter{footnote}{0}}  

\catcode`@=12
\relax

\def\figcap{\section*{Figure Captions\markboth
	{FIGURECAPTIONS}{FIGURECAPTIONS}}\list
	{Figure \arabic{enumi}:\hfill}{\settowidth\labelwidth{Figure
999:}
	\leftmargin\labelwidth
	\advance\leftmargin\labelsep\usecounter{enumi}}}
 \relax
\def\tablecap{\section*{Table Captions\markboth
	{TABLECAPTIONS}{TABLECAPTIONS}}\list
	{Table \arabic{enumi}:\hfill}{\settowidth\labelwidth{Table
999:}
	\leftmargin\labelwidth
	\advance\leftmargin\labelsep\usecounter{enumi}}}
 \relax
\def\reflist{\section*{References\markboth
	{REFLIST}{REFLIST}}\list
	{[\arabic{enumi}]\hfill}{\settowidth\labelwidth{[999]}
	\leftmargin\labelwidth
	\advance\leftmargin\labelsep\usecounter{enumi}}}
 \relax
\makeatletter
\newcounter{pubctr}
\def\publist{\@ifnextchar[{\@publist}{\@@publist}}
\def\@publist[#1]{\list
	{[\arabic{pubctr}]\hfill}{\settowidth\labelwidth{[999]}
	\leftmargin\labelwidth
	\advance\leftmargin\labelsep
	\@nmbrlisttrue\def\@listctr{pubctr}
	\setcounter{pubctr}{#1}\addtocounter{pubctr}{-1}}}
\def\@@publist{\list
	{[\arabic{pubctr}]\hfill}{\settowidth\labelwidth{[999]}
	\leftmargin\labelwidth
	\advance\leftmargin\labelsep
	\@nmbrlisttrue\def\@listctr{pubctr}}}
 \relax
\makeatother
\newskip\humongous \humongous=0pt plus 1000pt minus 1000pt

\newif\ifdtup

\relax

\begin{document}
\begin{titlepage}
\begin{center}
\vskip .5cm

\hfill CERN--TH/95--14\\
\hfill hep-th/9502007\\

\vskip .6cm

{\large \bf Anisotropic space-times in \\
 homogeneous string cosmology}

\vskip .5cm

{\bf  Nikolaos A. Batakis}
\footnote{Permanent address: Department of Physics, University of
Ioannina,
GR--45100
 Ioannina, Greece}
\footnote{e--mail address: batakis@surya20.cern.ch}

{\em Theory Division, CERN\\
     CH--1211 Geneva 23, Switzerland}\\

\vskip .2cm

and
\vskip .2cm

{\bf Alexandros A. Kehagias }
\footnote{e-mail address: kehagias@si.kun.nl}

{\em Phys. Dept., National Technical Univ.\\
GR--15773 Zografou Athens, Greece}
\end{center}

\vskip .5cm

\begin{center} {\bf ABSTRACT } \end{center}
\begin{quotation}\noindent
\end{quotation}
The dynamics of the early universe may have been profoundly influenced by
spatial anisotropies.
 A search for such backgrounds
in the context of string cosmology has uncovered
the existence of an entire class of (spatially) homogeneous
but not necessarily isotropic space-times, analogous to  the class  of
Bianchi-types  in general relativity. Configurations
with vanishing cosmological constant but  with non-vanishing
dilaton and antisymmetric field are explicitly found for
all types. This is a new class of  solutions, whose  isotropy
limits reproduce all known and, further, all possible FRW-type of models
in the string-cosmology context considered. There is always
an initial singularity and no inflation.
Other features of the general solutions, including their behaviour
\vskip.3in
\noindent
CERN--TH/95--14 \\
January 1995\\
\end{titlepage}
\newpage
\section{Introduction}

There are several compelling reasons (to be reviewed shortly) which indicate
that the dynamics of the early universe may have been
profoundly influenced by the presence of spatial
anisotropies just below the Planck or string scale \cite{1}.
It follows that  the  assumption
of a Friedmann-Robertson-Walker (FRW) behaviour may be hindering
important aspects of the early dynamics
and other cosmological issues
traceable to that era.  Thus  motivated we have relaxed the requirment of
spatial isotropy in the FRW-type of symmetry in order to study (spatially)
homogeneous
models as background space-times for string theory \cite{2},\cite{2'}.
We have found that  the one-loop beta-function
equations in four dimensions admit an entire class  of solutions
representing  homogeneous but not necessarily
isotropic space-times.
This class, which
will be
explicitly
presented in this paper,  may be considered as the string-cosmology
counterpart of what is known in general relativity as  the class
of vacuum Bianchi-type cosmologies
\cite{3},\cite{4}. Without being
excessively symmetric (they possess three Killing vectors compared to the six
of the FRW and the ten of the Minkowski space-time), the Bianchi-type
space-times allow for their full classification and a rather elegant
mathematical treatment. Even more important seems the fact that
any given  FRW behaviour may be viewed as the isotropic
limit of some homogeneous spacetime. It follows
that the study of such spacetimes (in the context of what
one might appropriately call `homogeneous string cosmology')
may offer us
tractable and promising models for  understanding  the dynamics of
anisotropy and its potentially crucial r$\hat{o}$le in the early universe,
until the attainment of the observed state of isotropy.

To establish notation we recall that the one-loop beta-function equations
for conformal invariance are \cite{5}
\begin{eqnarray}
R_{\mu\nu}-\frac{1}{4}H_{\mu\nu}^2-\nabla_\mu\nabla_\nu\phi&=&0,
\label{b1}\\
\nabla^2(e^\phi H_{\mu\nu\lambda})&=&0, \label{b2} \\
-R+\frac{1}{12}H^2+2\nabla^2
\phi+(\partial_\mu\phi)^2+\Lambda&=&0,\label{b3}
\end{eqnarray}
where the contractions $H_{\mu\nu}^2=H_{\mu\kappa\lambda}
{H_{\nu}}^{\kappa\lambda}
\, , H^2=H_{\mu\nu\lambda}H^{\mu\nu\lambda}$ involve the totally antisymmetric
field strenght $H_{\mu\nu\lambda}$, defined in terms of the potential
$B_{\mu\nu}$ as
\begin{equation}
H_{\mu\nu\rho}=\partial_\mu
B_{\nu\rho}+\partial_\rho B_{\mu\nu}+\partial_{\nu} B_{\rho\mu} \label{H}.
\end{equation}

In a 4-dimensional context, the set
(\ref{b1}--\ref{b3}) can be derived from an effective action
of the form
\begin{equation}
S_{eff}=\int d^4x
\sqrt{-g}e^{\phi}(R-\frac{1}{12}
H_{\mu\nu\rho}H^{\mu\nu\rho}+\partial_{\mu}\phi\partial^{\mu}\phi-\Lambda)
\label{b4'}.
\end{equation}
The constant $\Lambda$,   directly related to the
charge deficit $\delta c$ in the original theory,  is proportional to
$D-26$ or $D-10$ in the bosonic or heterotic string respectively.
It therefore vanishes at critical dimensions or
otherwise may be neglected at sufficiently
large curvatures $R$ or kinetic energies $(\nabla\phi)^2,\, H^2$ for
the dilaton and $B$ field. The action (\ref{b4'}),
already in the so-called sigma frame, may be recast as
 \begin{equation}
S_{eff}=\int d^4x
\sqrt{-\tilde{g}}(\tilde{R}-\frac{1}{12}
H_{\mu\nu\rho}H^{\mu\nu\rho}e^{2\phi}-\frac{1}{2}
\partial_{\mu}\phi\partial^{\mu}\phi-\Lambda e^{-\phi})
\label{b4}.
\end{equation}
wherein the metric is conformally related to the previous one by
\begin{equation}
\tilde{g}_{\mu\nu}=e^{\phi}g_{\mu\nu}\label{Ef}
\end{equation}
and it is being utilized to raise indices etc
in this more conventional `Einstein frame'.
Eq. (\ref{b1}) resumes then a
typical general relativistic form as
\begin{equation}
\tilde{R}_{\mu\nu}-\frac{1}{2}\tilde{R}\tilde{g}_{\mu\nu}=
\kappa^2(T_{\mu\nu}^{(\phi)}+
T_{\mu\nu}^{(H)}) \label{bx}
\end{equation}
with
\begin{eqnarray}
\kappa^2T_{\mu\nu}^{(\phi)}&=&\frac{1}{2}(\partial_\mu\phi
\partial_\nu\phi-\frac{1}{2}(\partial\phi)^2\tilde{g}_{\mu\nu}
-\Lambda e^{-\phi}\tilde{g}_{\mu\nu} )\, , \nonumber\\
\kappa^2T_{\mu\nu}^{(H)}&=&\frac{1}{4}e^{2\phi}
(H_{\mu\kappa\lambda}{H_\nu}^{\kappa\lambda}
-\frac{1}{6}H^2\tilde{g}_{\mu\nu})\, , \label{en}
\end{eqnarray}
the energy-momentum tensor for $\phi$ and $H$, while the remaining Eqs.
(\ref{b2},
\ref{b3}) become
\begin{eqnarray}
\tilde{\nabla}^2(e^\phi H_{\mu\nu\lambda})&=&0,\nonumber  \\
-\frac{1}{6}e^{2\phi}H^2
+\tilde{\nabla}^2\phi+\Lambda e^{-\phi}&=&0.\label{b8}
\end{eqnarray}

{}From experience in general relativity we know that searching for
the general solution to the system of Eqs. (\ref{b1}--\ref{b3}) or
equivalently  (\ref{bx}--\ref{b8}) is a hopeless and perhaps
meaningless enterprise. Without any underlying symmetry \cite{5'},\cite{KA} or
other
physical input, such a coupled system of non-linear differential
equations can supply an immence plethora of configurations
representing chaotic or other intractable excitations, which would have little
or nothing to do with the real world. On the other hand,
several interesting cosmological
solutions have been constructed explicitly
 either by directly solving the background field equations
 or by exploiting conformal
 theory techniques \cite{2},\cite{2'}, \cite{6}--\cite{11}.
 As far as we are  aware,  a
classification  of solutions
has yet to appear in the literature.

The structure of the rest of this paper is as follows.
In the next section we proceed with an introduction to homogeneous string
cosmology and the reasons which motivate its study. In section 3 we write down
the background field equations in the sigma frame, to be followed by explicit
solutions for each Bianchi type, presented in section 4. These reults are
briefly discussed in the last section and summarized in a Table. The latter
is preceeded by an Appendix with definitions and proceedures necessary for the
reproduction
of our results.

\section{Homogeneous string cosmology}

By relaxing the requirement of isotropy made in the FRW-type of models,
one is led to the study of   (spatially) homogeneous  but not necessarily
isotropic space-times $M^4$ \cite{1},\cite{KT},\cite{RS}.
The arguments which motivate this study
may be summarized as follows.
\vspace{.2cm}

1. The generally claimed isotropy of the universe
is deduced from fairly reliable observations and measurments, including
those on the cosmic  microwave background, which are  by far the most
accurate ones in cosmology.  However, all these results
have  been established for times
subsequent to the era at which the universe became transparent to
the electromagnetic radiation. Their extrapolation to earlier times
and, in particular, near the Planck or string scale is totally
unfounded.
\vspace{.2cm}

2. Statistical flactuation in the FRW models or otherwise small
perturbations of the initial conditions either grow or otherwise
do not smooth-out sufficiently fast to conform with the observed
state of the universe, thus giving rise to a host of
paradoxes such as the flatness and the horizon problems. It follows
that in any scenario (such as inflation) to resolve these problems,
the observed FRW behaviour should emerge as a consequence rather than
taken as starting  assumption.
\vspace{.2cm}

3. The kinetic and potential energy of the gravitational field
which may be attributed to the presence of anisotropy could
reach the same order of magnitute or even exeed the energy
attributed to any other field present in the effective action (\ref{b4})
at some time in the early evolution. Hence, the dynamics of the universe
at that time could be profoundly distorted if one neglected that
contribution.
\vspace{.2cm}

4. The presence and significance
of an initial singularity is hardly questioned
as  based on both observational
as well as theoretical grounds. However,  the dynamics and structure
of such an initial state as predicted by the FRW behaviour may
change profoundly in the presence of anisotropy.
\vspace{.2cm}

As in conventional general relativity, homogeneous
backgrounds in string cosmology   may be
defined as  those 4-dimensional space-time
manifolds which admit an r-parameter group of isometries $G_r$ whose
orbits in $M^4$ are 3-dimensional space-like hypersurfaces $\Sigma^3$.
The latter are precisely the hypersurfaces of homogeneity  on
which a $G_3$ subgroup of $G_r$ acts transitively. Clearly,
 $3\leq r\leq 6$, so that any
remaining symmetry ( and the corresponding independent Killing vectors)
must generate the (r-3)-dimensional
isotropy subgroup of $G_r$. Since there is no
two-dimensional rotation group, (r-3) can only have the values $0,1,3$,
with the last one associated with the maximal FRW-type of symmetry. The
action of $G_3$ is almost always simply transitive on its orbits.
The exception is
been realized in the Kantowski-Sachs type of metric, in which $G_3$
is actually acting on 2-dimensional spacelike surfaces  of maximal symmetry.
In the typical case which,  as mentioned,  involves
 a simply-transitive $G_3$, the most
general metric may be written as
\begin{equation}
ds^2=-dt^2+g_{0i}dt\sigma^i+g_{ij}\sigma^i\sigma^j, \label{met}
\end{equation}
where the metric coefficients may be functions of the time t only
and $\{\sigma^i,\, i=1,2,3\}$ is a basis of 1-forms,
invariant under the left action of $G_3$.

Our study of Bianchi-type models
in string cosmology will be made under the assumption of a vanishing
cosmological constant. We will also restrict ourselves to metrics
which are diagonal in the non-holonomic frame employed in Eq. (\ref{met})
namely  are expressible as
\begin{equation}
ds^2=-dt^2+a_1(t)^2(\sigma^1)^2+a_2(t)^2(\sigma^2)^2+a_3(t)^2(\sigma^3)^2
 \label{met1}
\end{equation}
All  possible isometry  groups $G_3$ of  the metric (\ref{met1})
are known and have been fully classified.
Each  one of the existing nine Bianchi types is identified
by the corresponding set of group-structure constants $C^i_{jk}$ which define
the relation
\begin{equation}
d\sigma^i=-\frac{1}{2}C^i_{jk}\sigma^j\wedge\sigma^k
\end{equation}
or, equivalently,
\begin{equation}
[\xi_i,\xi_j]=-C^k_{ij}\xi_k,
\end{equation}
where the set of the three independent Killing vectors ${\xi_i}$  forms a
basis dual to $\{\sigma^i\}$.

The simplest case is that of a Bianchi-type I model in which $G_3$ is just the
abelian tranlation group $T_3$ and the hypersurfaces of homogeneity $\Sigma^3$
are Euclidean  3-dimensional sheets.
At the other end of the  Bianchi classification we find the type-IX
models for $G_3=SO(3)$,  and
 $\Sigma^3$ sections with $S^3$ topology. We also note that
Bianchi-type $VI_h$ space-times involve an entire one-parameter family of $G_3$
 groups, parametrized by $h$ which takes all real values exept 0,1.
Likewise, a  one-parameter family of $G_3$ groups
is involved in  the  Bianchi-type $VII_h$ case,
here with $-2<h<2$. The situation is much simpler with the FRW models,
which of course are fully classified by $k=0,-1,+1$
and also have an $SO(3)$-isotropy
group in addition to their $G_3$. All these $G_6$ space-times are  typically
 realized by Bianchi-types $I$ (and $VII_0$), $V$,$IX$
respectively.
 All Bianchi-types (as summarized in the Table) have been  further classified
as being of class A if their
adjoint representation  is traceless, otherwise they are of class B.

\section{The background field equations}
In the next section we will examine   each Bianchi type  separately.
However, before doing that it will be usefull to  present in this
section  features of the background field equations
 which are common to all types.
\subsection{The dilaton field and the totally antisymetric field strength}
To respect homogeneity, any scalar such as the dilaton $\phi$
 must be a constant on each hypersurface
$\Sigma_3$ so that, in $M^4$,  $\phi$ can at most be a function of the time t.
 On similar grounds we may express the 3-form $H$ as
\begin{equation}
H=A\sigma^1\wedge\sigma^2\wedge\sigma^3, \label{h}
\end{equation}
which, with $A$ considered as at most a function
of t only, is equivalent to the usual trading of $H$
for a time dependent axion field in 4D.
 Here however, in view of the $dH=0$ requirment,  $A$ must  be a constant
not only on $\Sigma^3$ but everywhere in $M^4$. It is reminded however that
in conventional (namely holonomic) coordinates, the components of the same $H$
will
certainly depend on t and, in general, on the spatial coordinates as well. It
follows
that Eq. (\ref{b2}) is then automatically satisfied for any
time-dependent configuration of the dilaton field.  With a
dot standing for $d/dt$, Eq. (\ref{b3}) may now be expressed as
\begin{equation}
\ddot{\phi}+(\dot{\phi})^2+3\frac{\dot{a}}{a}\dot{\phi}+
\frac{1}{6}(\frac{A}{a^3})^2=0. \label{f}
\end{equation}
where we have introduced the radius $a$ defined by
\begin{equation}
a^3=a_1a_2a_3, \label{a}
\end{equation}
 so that $\dot{a}/a$ is the mean Hubble constant of any comoving
volume element (the latter is always proportional to $a^3$).
In terms of the new time coordinate $\tau$ defined by
\begin{equation}
d\tau=a^{-3}e^{-\phi} dt, \label{t}
\end{equation}
 and  with a prime standing for $d/d\tau$ we may re-express Eq. (\ref{f})
as
\begin{equation}
\phi^{\prime\prime}+A^2e^{2\phi}=0.
\end{equation}
The  general solution of the above equation
may be expressed as
\begin{equation}
e^{-\phi}=\cosh(N\tau) +\sqrt{1-\frac{A^2}{N^2}}\sinh(N\tau),
\label{sol}
\end{equation}
where $N$ is a constant, while  a second
constant of integration has been absorbed to fix the origin of $\phi$.
In an orthonormal frame $\{\omega^\mu\}$ (cf. Appendix), the energy-momentum
tensors
 (\ref{en}) for the $\phi$ and
$H$ fields  have the form of
\begin{eqnarray}
\kappa^2T_{\mu\nu}^{(\phi)}&=
&\frac{1}{4}a^{-6}e^{-3\phi}(\frac{d\phi}{d\tau})^2
diag(1,1,1,1), \nonumber \\
\kappa^2T_{\mu\nu}^{(H)}&=&\frac{1}{4}A^2a^{-6}
diag(1,1,1,1). \label{ene}
\end{eqnarray}
These expressions will be briefly recalled in the last section.

\subsection{The gravitational field equations}
As outlined in the Appendix, the set of Eqs. (\ref{b1}) involves
three second order differential equations for the radii $a_i(t)$
which are of the form
\begin{equation}
(\ln a_i^2e^{\phi})^{\prime\prime}+2V_i=0,\,\,(i=1,2,3) \label{aa}
\end{equation}
Although in the sigma-model frame, these equations are expressed more concisely
in terms of the variables $a_i^2e^{\phi}$, which, according to (\ref{Ef}), are
just
the metric coefficients $\tilde{a}_i^2$ in the Einstein frame.
Each `potential' $V_i$ is generally dependent on all three radii $\tilde{a}_j$.
 Eqs. (\ref{aa}) are in fact the $(ii)$ components of the set (\ref{b1}).
Their solutions $a_i(t)$ are subject to

i) The initial value equation
\begin{equation}
\sum_{i<j}(\ln a_i^2e^{\phi})^{\prime}
(\ln a_j^2e^{\phi})^{\prime}+ \\
2\sum V_i=A^2 , \label{in}
\end{equation}
which is essentialy the $(00)$ equation in the set (\ref{b1}).

ii) A set of constraint equations which are of the form
\begin{equation}
R_{\mu\nu}=0, \, \, (\mu\neq\nu).  \label{cons}
\end{equation}
and are due to the emergence of non-vanishing off-diagonal components
of the Ricci tensor. For the type of metrics considered
here, these arise only in the case of class-B types.

\section{Explicit solutions}

In this section we examine the background field equations and solve them
for each Bianchi-type separately. We recall
that the expressions (\ref{sol},\ref{h}) which specify the dilaton and the
$H$-field hold invariably for all Bianchi types.
The results presented here
will be further discussed in the last section.
\subsection{Type I}
As mentioned already,  in  Type-I models  the $\Sigma^3$ hypersurfaces
are 3-dimensional flat Euclidean sheets generated by the group
of  abelian  translations $T_3$. The set (\ref{aa}) is simply
\begin{equation}
(\ln a_i^2e^{\phi})^{\prime\prime}=0,
\end{equation}
whose general solution is
\begin{equation}
a_i^2e^{\phi}=L_ie^{p_i\tau}. \label{I}
\end{equation}
There are no constraint equations but the initial value equation (\ref{in})
imposes on the constants $p_i$  the Kasner-like restriction
\begin{equation}
\sum_{i<j}p_ip_j=A^2. \label{p}
\end{equation}
 When $A=0$ the above solution reduces to the one  found in \cite{2}.
When the isometry group is enlarged to $T_3\!\times\!SO(3)$,
namely at  the isotropic limit attained at $a_1=a_2=a_3$,
the above solution reduces to the one given
in \cite{8}.

\subsection{Type II}
Here  Eqs. (\ref{aa})
are expressed as
\begin{eqnarray}
(\ln a_1^2e^{\phi})^{\prime\prime}+a_1^4e^{2\phi}&=&0, \nonumber \\
(\ln a_2^2e^{\phi})^{\prime\prime}-a_1^4e^{2\phi}&=&0, \nonumber \\
(\ln a_3^2e^{\phi})^{\prime\prime}-a_1^4e^{2\phi}&=&0,
\end{eqnarray}
which admit the general solution
\begin{eqnarray}
a_1^2e^{\phi}&=&\frac{p_1}{\cosh(p_1\tau)}, \nonumber \\
a_2^2e^{\phi}&=&L_2^2\cosh(p_1\tau)e^{p_2\tau}, \nonumber \\
a_3^2e^{\phi}&=&L_3^2\cosh(p_1\tau)e^{p_3\tau}. \label{II}
\end{eqnarray}
There are no constraint equations but Eq. (\ref{in}) subjects the
constants $p_i$ to the restriction
\begin{equation}
p_2p_3-p_1^2=A^2.
\end{equation}
\subsection{Type III}
This is a class B space-time subject to the constraint equation
\begin{equation}
(\ln \frac{a_3}{a_1})^{\prime}=0
\end{equation}
The set of Eqs. (\ref{aa}) can be written as
\begin{eqnarray}
(\ln a_1^2e^{\phi})^{\prime\prime}-2(a_1a_2e^{\phi})^2&=&0, \nonumber \\
(\ln a_2^2e^{\phi})^{\prime\prime}&=&0,
\end{eqnarray}
where without loss of generallity we have assumed that $a_1=a_3$.
The general solution to the above system is
\begin{eqnarray}
a_1^2e^{\phi}&=&\frac{p_1}{\sinh^2(p_1\tau)}e^{p_2\tau}, \nonumber \\
a_2^2e^{\phi}&=&p_1e^{-p_2\tau}, \label{III}
\end{eqnarray}
with the integration constants subject to the
restriction
\begin{equation}
4p_1^2-p_2^2=A^2
\end{equation}
imposed by Eq. (\ref{in}).
\subsection{Type IV}
In this (class B) case the constraint equations  (\ref{cons}) give
\begin{eqnarray}
\frac{a_1}{a_2a_3^2}&=&0 , \nonumber \\
\frac{1}{a_3}(\ln \frac{a_1a_2}{a_3^2})^{\prime}&=&0
\end{eqnarray}
so that, within the present context, all solutions are singular everywhere in
$M^4$.
\subsection{Type V}
The constraint equations  (\ref{cons})
impose in this case the restriction  ($q_1$ is a constant)
\begin{equation}
a_1^2=q_1a_2a_3.
\end{equation}
The field equations (\ref{aa}) are
\begin{eqnarray}
(\ln a_1^2e^{\phi})^{\prime\prime}-4(a_2a_3e^{\phi})^2&=&0, \nonumber \\
(\ln a_2^2e^{\phi})^{\prime\prime}-4(a_2a_3e^{\phi})^2&=&0, \nonumber \\
(\ln a_3^2e^{\phi})^{\prime\prime}-4(a_2a_3e^{\phi})^2&=&0.
\end{eqnarray}
It follows  that the most general solution which
satisfies the above constraint is
\begin{eqnarray}
a_1^2e^{\phi}&=&\frac{q_1p_1}{2\sinh(p_1\tau)}, \nonumber \\
a_2^2e^{\phi}&=&\frac{q_2p_1}{2\sinh(p_1\tau)}e^{p_2\tau}, \nonumber \\
a_3^2e^{\phi}&=&\frac{p_1}{2q_2\sinh(p_1\tau)}e^{-p_2\tau}. \label{V}
\end{eqnarray}
 The constants $p_1,p_2$ are subject to the restriction
\begin{eqnarray}
3p_1^2-p_2^2=A^2
\end{eqnarray}
imposed by Eq. (\ref{in}). When the isometry group is enlarged to
a $G_6$ the above solution reduces to its    isotropic limit
attained at $p_1=A/\sqrt{3}$ and $q_1=q_2=1$. This represents an
open (k=-1) FRW model with radius
\begin{eqnarray}
a^2e^{\phi}=\frac{A}{2\sqrt{3}\sinh(\frac{A}{\sqrt{3}}\tau)},
\end{eqnarray}
which has been already discussed in the quoted literature.

\subsection{Type $VI_h$}
The $G_3$ involved in this type of models is actually a
continuous 1-parameter family of groups parametrized by  $h$,
with the values  $h\neq 0,1$ typically excluded as giving rise to
Bianchi types III and V respectively. To facilitate calculations
we will  assume for the moment  that we also have
$h\neq -1$ and present the Bianchi-type
$VI_{-1}$ space-time separately in the next subsection.
 These models are subject by Eqs. (\ref{cons})
to the constraint
\begin{equation}
\frac{a_2^ha_3}{a_1^{h+1}}=\frac{q_2^{h-1}}{q_1^{h+1}},\label{const}
\end{equation}
where the rhs has been expressed in terms of the numerical constants $q_1,q_2$,
introduced for later convenience.
The field equations  (\ref{aa}) are
\begin{eqnarray}
(\ln a_1^2e^{\phi})^{\prime\prime}-2(h^2+1)
(a_2a_3e^{\phi})^2&=&0, \nonumber \\
(\ln a_2^2e^{\phi})^{\prime\prime}-2h(h+1)
(a_2a_3e^{\phi})^2&=&0, \nonumber \\
(\ln a_3^2e^{\phi})^{\prime\prime}-2(h+1)(a_2a_3e^{\phi})^2&=&0.
\label{qqq}
\end{eqnarray}
This system can be fully integrated
and the most general solution which satisfies the above constraint is
given by
\begin{eqnarray}
a_1^2e^{\phi}&=&q_1^2(\frac{p_1}{h+1})^{\frac{2(h^2+1)}{(h+1)^2}}
\sinh(p_1\tau)^{-\frac{2(h^2+1)}{(h+1)^2}}
e^{\frac{(h-1)}{h+1}p_2\tau}, \nonumber \\
a_2^2e^{\phi}&=&q_2^2(\frac{p_1}{h+1})^{\frac{2h}{h+1}}
\sinh(p_1\tau)^{-\frac{2h}{h+1}}
e^{p_2\tau}, \nonumber \\
a_3^2e^{\phi}&=&q_2^{-2}(\frac{p_1}{h+1})^{\frac{2}{h+1}}
\sinh(p_1\tau)^{-\frac{2}{h+1}}
e^{-p_2\tau}. \label{VI}
\end{eqnarray}
The constants $p_1,p_2$ are  subject to the restriction
\begin{equation}
\frac{4(h^2+h+1)}{(h+1)^2}p_1^2-p_2^2=A^2, \label{VIiso}
\end{equation}
as required by Eq. (\ref{in}). We observe however that, in this particular
type,
the restriction imposed by the initial value equation can be evaded.
In other words, the constants $p_1,p_2$ may be chosen arbitrarily,
but at the expense of fine-tunning for specifically chosen group parameters
$h$,
so that Eq.(\ref{VIiso}) is satisfied.

\subsection{Type $VI_{-1}$}
For $h=-1$ the constraint equation (\ref{const}) may without loss
of generality be written as  $a_2=a_3$ so that  Eqs.(\ref{qqq}) reduce to
\begin{eqnarray}
(\ln a_1^2e^{\phi})^{\prime\prime}-4
a_2^4e^{2\phi}&=&0, \nonumber \\
(\ln a_2^2e^{\phi})^{\prime\prime}&=&0
\end{eqnarray}
The general solution to this system is
\begin{eqnarray}
a_1^2e^{\phi}&=& q_1^2p_1 exp(q_2^4e^{2p_2\tau})e^{p_1\tau}, \nonumber \\
a_2^2e^{\phi} &=& q_2^2p_2 e^{p_2\tau}, \label{VI-1}
\end{eqnarray}
with the constants $p_1,p_2$  subject to the restriction
\[
p_2^2+2p_1p_2=A^2,
\]
coming from Eq. (\ref{in}).

\subsection{Type $VII_h$}
In this case the constraints imposed by (\ref{cons}) are
\begin{eqnarray}
\frac{h}{a_3}(\ln \frac{a_2}{a_3})^{\prime}&=&0, \nonumber \\
\frac{ha_2}{a_1a_3}&=&0 \label{ss}
\end{eqnarray}
We  observe that  all solutions are singular  everywhere  unless $h=0$.
In the latter case (which involves
space-times of  class A)  the field equations  (\ref{aa})
may be expressed as
\begin{eqnarray}
(\ln a_1^2e^{\phi})^{\prime\prime} +
(a_1^4-a_2^4)e^{2\phi}&=&0, \nonumber \\
(\ln a_2^2e^{\phi})^{\prime\prime} +
(a_2^4-a_1^4)e^{2\phi}&=&0, \nonumber \\
(\ln a_3^2e^{\phi})^{\prime\prime}-
(a_1^2-a_2^2)^2e^{2\phi}&=&0. \label{VII0}
\end{eqnarray}
We have not being able to write down
the general solution to the above equations in closed form.
We observe, however, that by enlarging the isotropy
group  to a $G_4$ (namely with
$a_1=a_2$) the above equations  reduce to  the Bianchi-type I
set so that the corresponding solutions
may be recovered from there.
\subsection{Type VIII}
For space-times in this type there are no constraint equations and
Eqs. (\ref{aa}) are written as
\begin{eqnarray}
(\ln a_1^2e^{\phi})^{\prime\prime} +
(a_1^4-(a_2^2+a_3^2)^2)e^{2\phi}&=&0, \nonumber \\
(\ln a_2^2e^{\phi})^{\prime\prime} +
(a_2^4-(a_3^2-a_1^2)^2)e^{2\phi}&=&0, \nonumber \\
(\ln a_3^2e^{\phi})^{\prime\prime} +
(a_3^4-(a_1^2+a_2^2)^2)e^{2\phi}&=&0,
\end{eqnarray}
A general analytic solution to the above system seems attainable only
after enlargment of the isotropy group to a $G_4$
(e.g., with $a_1=a_3$). We then have
\begin{eqnarray}
a_1^2e^{\phi}=a_3^2e^{\phi}&=&\frac{p_1^2
\cosh(p_2(\tau-\tau_0))}
{p_2\sinh^2(p_1\tau))}, \nonumber \\
a_2^2e^{\phi}&=& \frac{p_2}{\cosh(p_2(\tau-\tau_0)}, \label{VIII}
\end{eqnarray}
with the constants subject to the restriction
\[
4p_1^2-p_2^2=A^2
\]
as required by Eq. (\ref{in}).

\subsection{Type IX}
Space-times of  this type  have been extensively studied
in general relativity and they include
the well-known Taub and mixmaster models \cite{3},\cite{3'}.
No analytic solution has been given for the latter case.  In the former
case the solution is obtained after
the enlargement  of
the original isometry group to $SO(3)\!\times\!U(1)$.
In the present context the field equations
 (\ref{aa}) are written in the general case as
\begin{eqnarray}
(\ln a_i^2e^{\phi})^{\prime\prime} +
(a_i^4-(a_j^2-a_k^2)^2)e^{2\phi}=0, \label{mix}
\end{eqnarray}
with $(ijk)$ taken cyclically as $(123)$ and there are no constraint equations.
Eqs. (\ref{mix}) are precisely those of the mixmaster system which is
known not to yield to an analytic treatment.
Nevertheless, as in type VIII, under
enlargment of the isometry to a $G_4$ (e.g., with $a_1=a_3$) the
system of equations (\ref{mix})  can be  fully integrated.
We have  thus reproduced
what may be considered as the string-cosmology analog of the
Taub universe.
This result may be expressed as
\begin{eqnarray}
a_1^2e^{\phi}=a_3^2e^{\phi}&=&\frac{p_1^2
\cosh(p_2(\tau-\tau_0))}
{p_2\cosh^2(p_1\tau)}, \nonumber \\
a_2^2e^{\phi}&=& \frac{p_2}{\cosh(p_2(\tau-\tau_0)}, \label{IX}
\end{eqnarray}
where  the constants  $p_1,\,p_2$ are subject to the restriction
\[
4p_1^2-p_2^2=A^2
\]
as required by Eq. (\ref{in}). Our solution reproduces
Taub's metric
if we switch off the $H$ field ($A=0$).
It is interesting to note that, contrary to Taub's metric, the present
generalization possesses a limit of complete isotropy, attainable at
$p_1=p_2=A/\sqrt{3}$.
We then obtain a closed (k=1) FRW model
with radius
\begin{equation}
a^2e^{\phi}=\frac{A}{\sqrt{3}\cosh(\frac{A}{\sqrt{3}}\tau)}, \label{IXO}
\end{equation}
which has already been discussed in the quoted literature.

\section{Conclusions}
We have explicitly found a general class of
background space-times in the context of
homogeneous string cosmology,
with non-vanishing dilaton and antisymmetric H field.
Our findings have been obtained under the assumption of vanishing cosmological
constant and for
metrics which are diagonal in the invariant $\{\sigma^i\}$ basis.
It follows that (modulo the doubling of classes caused by a non-vanishing
cosmological constant)
there exists only one other general class in homogeneous string cosmology. It
is that of metrics (\ref{met}) which
cannot be diagonalized in the above ivariant basis. The distinction
between these two fundamental classes can be better illuminated on more
physical grounds. We observe
that, in the non-diagonal case, the gradient
of the axion field will generally depend not only on t but
on  spatial coordinates as well. Equivalently (and disregarding here some
marginal cases), the dual of H
would no longer be orthogonal to the hypersurfaces of homogeneity $\Sigma^3$.
One may now deduce that these `tilted' models represent rotating universes
\cite{1},\cite{4},\cite{RS}.
In the respective space-times, cosmic vorticity
will emerge in addition to
expansion and shear,
already existing in the case of diagonal metrics to which we now return.
Explicit expressions for the standard parameters of expansion
(Hubble constants) and shear (anisotropy) for all Bianchi types
can be red-off (or easily found) directly from our results in section 4.
All general solutions presented in that section are new and reduce to known
results and isotropic limits as
summarized in the Table. Following the reasoning in section 2, it may be
expected that the
respective models could illuminate the dynamics of the early universe,
where string theory probably has its best (if not the only)
chance of been confronted with reality.

The following may be noted as  preliminaries to a more detailed study
(e.g., in the approach utilized in \cite{7} for the isotropic case).
We firstly recall that
our results may be equivalently expressed
in the Einstein frame (now aptly in terms of the radii $\tilde{a_i}$ etc.)
and examined there in a general relativistic
context. We already have at out disposal
the energy momentum tensors (\ref{ene})
for  the $\phi$ and
H fields which
have the form of the energy-momentum tensor of a
perfect fluid. Such an identification assigns a `stiff matter' behaviour
to these fields, while the associated
$\rho=p$ equation of state does not upset the
energy conditions.The former result reveals a gravity-like nature for the
dilaton and H field (of importance when generating these fields
from vacuum configurations), while the latter makes inevitable
the presence of an initial singularity (which is indeed the case with all our
solutions).
On similar grounds one may generally verify that none of our solutions
can have inflation. This follows from the sign of the quantity
$\frac{d^2\tilde{a}}{d\tilde{t}^2}$, which is always negative
- thus forbidding superluminal expansion.
Also of apparent theoretical (and possibly
observational as well) significance  is the inter-relation between
the Hubble constants along the principal directions of anisotropy
and the magnitute of the H-field strength  as required by
the initial value equation (\ref{in}). A particular example of this
inter-relation
has been the existence of the $SO(3)$ isotropy limit in our generalised Taub
metric (\ref{IX}). Indeed, at the $A=0$ limit, the isotropic solution
(\ref{IXO}) collapses to a single singular point, in
agreement with the non-existence of a full-isotropy limit in Taub's solution.

Obviously there  are several further issues to be
investigated in the more general context of
string theory.
We will only examine below   whether new solutions
can be generated from the above by   duality transformations.
As well known, target space  duality is
one of the most fundamental symmetries in string
theory  relating, as it does, small to large scales
(roughly speaking a radius $a_0$ in the original metric
 to its inverse  $a_0^{-1}$ in the dual) \cite{D},\cite{D'}.
The  calculations for abelian duality are straightforward but rather awkward to
be  explicitly
presented here, the reason being that
they had to be carried-out in holonomic coordinates, to which all
our solutions were firstly re-expressed. Fortunately however, the final result
(summarized in the Table) may be expressed
quite  concisely, at least for certain Bianchi types. These may be
grouped as  $(I,II)$ and
$(VI_h)$ (with the h-parameter typically allowed hereafter to also take the
values
0,1 for types $III,V$ respectively).
These types retain their homogeneity and in fact their  duals
reproduce  metrics as those in the
 Bianchi-types   $(II,I)$ and $(VI_{-h})$
respectively in that order. The situation is considerably more complex for
types
$IV,VII_h,VIII,IX$
for which homogeneity
seems to be generally lost under duality. However, when equipped with a
$G_4$ symmetry, these spacetimes essentially reproduce themselves.
In particular, this is precisely the case with our generalized-Taub
solution which reproduces duals with the same type-IX metric.

\vspace{1cm}

We would like to thank I. Bakas for discussions.

\vspace{1cm}

{\it NOTE ADDED}:

Ref. \cite{Gas}, which appeared
shortly after the completion of the present work
addresses part of the same general problem.   The Bianchi types examined there
correspond to our
types $I,\,II$ and $VI_{h}$ for vanishing H-field.

\newpage

\appendix{\large{\bf Appendix}}
\vspace{.5cm}

Here we review the formalism and definitions needed for the reproduction
of our results. Computations
simplify considerably if
carried out directly in the  invariant basis $\{\sigma^i\}$
rather than in conventional holonomic coordinates.
For even further simplification one may  express
(\ref{met1}) as a Minkowski metric, namely as
\[
ds^2=\eta_{\mu\nu}\omega^\mu\omega^\nu
\]
with $\eta_{\mu\nu}=diag(-1,1,1,1)$. In this orthonormal  non-holonomic
expression, which is defined in the sigma frame, the $\{\omega^\mu\}$ basis has
been chosen as
\begin{eqnarray}
\omega^0&=&dt, \nonumber \\
\omega^i&=&a_i(t)\sigma^i \,\, (no\,\, sum), \label{sig}
\end{eqnarray}
We now recall that, for vanishing torsion, Cartan's first structure equation
may be expressed as
\begin{equation}
d\omega^\mu=\gamma^\mu_{\nu\rho}\omega^\nu\wedge\omega^\rho. \label{first}
\end{equation}
The coefficients $\gamma^\mu_{\nu\rho}$ are the components of the
connection 1-form
$\omega^\mu_\nu$, so that
\[
\gamma_{\mu\nu\rho}=-\gamma_{\nu\mu\rho}.
\]
The components of the Riemann tensor are now supplied by Cartan's
second structure equation which may be here expressed as
\begin{equation}
\frac{1}{2}R^\lambda_{\mu\rho\nu}\omega^\rho\wedge\omega^\nu=
d\omega^\lambda_\mu+\omega^\lambda_\rho\wedge\omega^\rho_\mu.
\end{equation}
Hence, the components of the Ricci tensor may be expressed as
\begin{equation}
R_{\mu\nu}=R^\lambda_{\mu\lambda\nu}=
\partial_\rho \gamma^\rho_{\mu\nu}
-\partial_\nu \gamma^\rho_{\mu\rho} -
\gamma^\rho_{\mu\lambda}\gamma^\lambda_{\nu\rho}+
\gamma^\rho_{\lambda\rho}\gamma^\lambda_{\mu\nu}.
\end{equation}
It is precisely the above  components which have been utilised in writting
down the gravitational field equations in section 3.2.
\newpage
\leftmargin=0cm
\begin{center}
{\bf Bianchi-type  classification}
\end{center}

\begin{tabular}{|c|c|c|c|c|c|c|} \hline
Type&Cl.&$d\sigma^i=\frac{1}{2}C^i_{jk}\sigma^j\sigma^k $&
Coordinate basis& $a_i(t)$
& FRW&Dual \\
\hline \hline

 & & $d\sigma^1=0$& && \\
I&A&$d\sigma^2=0$&$ \sigma^i=dx^i$&(\ref{I})& YES& II\\
 & & $d\sigma^3=0$&  && \\ \hline
 & & $d\sigma^1=\sigma^2\wedge\sigma^3$&$\sigma^1=dx^2-x^1dx^3$&& & \\
II&A&$d\sigma^2=0$&$\sigma^2=dx^3$&(\ref{II})&NO&I\\
 & & $d\sigma^3=0$&$ \sigma^3=dx^1$& & &\\ \hline
 & & $d\sigma^1=0$&$\sigma^1=dx^1$& && \\
III&B&$d\sigma^2=0$&$\sigma^2=dx^2$&(\ref{III})&NO& III\\
 & & $d\sigma^3=\sigma^1\wedge\sigma^3$&$\sigma^3=e^{x^1}dx^3$&&  & \\ \hline
 & & $d\sigma^1=\sigma^1\wedge\sigma^3+ $&&&& \\&&
$\sigma^2\wedge\sigma^3$&
$\sigma^1=e^{-x^1}dx^2-x^1e^{-x^1}dx^3$&& &\\
IV&B&$d\sigma^2=\sigma^2\wedge\sigma^3$&$\sigma^2=e^{-x^1}dx^3$
&${}^\ast$&NO&${}^\ast$\\
 & & $d\sigma^3=0$&$\sigma^3=dx^1$& & & \\ \hline
 & & $d\sigma^1=0$&$\sigma^1=dx^1$& & &\\
V&B&$d\sigma^2=\sigma^1\wedge\sigma^2$&$\sigma^2=e^{x^1}dx^2$&(\ref{V})
&YES&$VI_{-1}$\\
 & & $d\sigma^3=\sigma^1\wedge\sigma^3$&$\sigma^3=e^{x^1}dx^3$&  & &\\ \hline
$VI_{h}$ & & $d\sigma^1=0$& $\sigma^1=dx^1$&(\ref{VI})&& \\
&B&$d\sigma^2=h\sigma^1\wedge\sigma^2$&$\sigma^2=
e^{hx^1}dx^2$&&NO &$VI_{-h}$\\ \cline{1-1}\cline{5-5}
$VI_{-1}$ & & $d\sigma^3=\sigma^1\wedge\sigma^3$&$\sigma^3=e^{x^1}dx^3$&
\ref{VI-1}& &\\ \hline
$VII_h$ &B & $d\sigma^1=-\sigma^2\wedge\sigma^3$
&$ {}^\star $
& ${}^\ast$&NO&${}^\ast$ \\
$h\neq 0$&&$d\sigma^2=\sigma^1\wedge\sigma^3+$
&$\sigma^1=(A-kB)dx^2-Bdx^3$&&&\\
\cline{1-1}\cline{5-7}&&$h\sigma^2\wedge\sigma^3$&
$\sigma^2=Bdx^2+(A+kB)dx^3$&${}^\ast{}^\ast$
&&\\
$VII_0$ &A & $d\sigma^3=0$&$ \sigma^3=dx^1$&(\ref{VII0}) &YES&I \\ \hline
 & & $d\sigma^1=\sigma^2\wedge\sigma^3$&
$\sigma^1=dx^1
+((x^1)^2-1)dx^2+$ &&& \\ &&& $(x^1+x^2-x^2(x^1)^2)dx^3$&
? &&? \\ \cline{5-7}
VIII&A&$d\sigma^2=-\sigma^3\wedge\sigma^1$&
$\sigma^2=dx^1
+((x^1)^2+1)dx^2+ $&&&\\&&& $(x^1-x^2-x^2(x^1)^2)dx^3$&
${}^\ast{}^\ast$&NO&\\
 & & $d\sigma^3=\sigma^1\wedge\sigma^2$&
$\sigma^3=2x^1dx^2+(1-2x^1x^2)dx^3$&(\ref{VIII})  &&VIII \\ \hline
 & & $d\sigma^1=\sigma^2\wedge\sigma^3$&
$\sigma^1=-\sin x^3dx^1+$ &&&\\ &&& $\sin x^1\cos x^3dx^2$&?&&?
\\ \cline{5-7}
IX&A&$d\sigma^2=\sigma^3\wedge\sigma^1$&
$\sigma^2=\cos x^3dx^1+$ &&& \\&&&$ \sin x^1 \sin x^3dx^2$&${}^\ast{}^\ast$
&YES&IX\\
 & & $d\sigma^3=\sigma^1\wedge\sigma^2$&$\sigma^3=\cos x^1dx^2+dx^3$&
(\ref{IX}) & \\ \hline
\end{tabular}

\noindent
${}^\ast\,$  Solutions singular everywhere.\\
${}^\ast {}^\ast$ Solution given for $a_1=a_3$. \\
 ${}^\star$ $  A=e^{-kx^1}\cos(qx^1),\, B=-\frac{1}{q}e^{-kx^1}
\sin(qx^1),\, \, \, \,
[k=\frac{h}{2}, \, q=\sqrt{1-k^2}]$


\newpage


\begin{thebibliography}{99}
\bibitem{1}  M.A.H. MacCallum, {\em ``Anisotropic and Inhomogeneous
Relativistic
Cosmologies"}, in {\em ``General Relativity-An Einstein Centenary Survey"},
eds.
S. W. Hawking and W. Israel, Cambridge Univ. Press, Campridge, 1979:\\
S. Barrow and D.H. Sonoda, Phys. Rep. 139 (1986)1.
\bibitem{2} M. Muller, Nucl. Phys. B337 (1990)37.
\bibitem{2'} E.J. Copeland, A. Lahiri and D. Wands, Phys. Rev. D50 (1994)4868;
{\em ``Srting cosmology with a time-dependent antisymmetric-tensor potential"}
preprint Susssex-Ast/94/10-1, hep/th/9410136.
\bibitem{3} A. Taub, Ann. Math. 53 (1951)472.
\bibitem{4} G.F.R. Ellis and M.A.H. MacCallum,
Commun. Math Phys. 12 (1969)108; 19 (1970)31.
\bibitem{5} E. Fradkin and A. Tseytlin, Nucl. Phys. B261 (1985)1;\\
C. Callan, D. Friedan, E. Martinec and M. Perry,
Nucl. Phys. B262 (1985)593.
\bibitem{5'}  I. Bakas, Nucl. Phys. B 428 (1994)374;
{\em ``Space-time interpretation of S-duality and supersymmetry violation of
T-duality"}, preprint CERN-th.7473/94, hep-th/9410104.
\bibitem{KA} N. Sanchez and G. Veneziano, Nucl. Phys. B333 (1990)253;\\
B.A. Campbell, A. Linde and K.A. Olive, Nucl. Phys. B355 (1991)146.
\bibitem{6}R. Myers, Phys. Lett. B199 (1987)371;\\
K.A. Meissner and G. Veneziano, Phys. Lett. B267(1991)33;\\A. Sen, Phys. Lett.
B271 (1991)295;\\M.Gasperini, J. Maharana and G. Veneziano, Phys. Lett. B272
(1991)277;\\ M. Gasperini and G. Veneziano, Phys. Lett. B277 (1992)256;
 Mod. Phys. Lett. A8 91993)3701; M. Gasperini, R. Ricci and
G. Veneziano, Phys. Lett. B319 (1993)438; \\
A.A. Tseytlin, Class Quantum Grav. 9 (1992)979.
\bibitem{7} R. Brandenberger and C. Vafa, Nucl. Phys. B316 (1988)319;\\
A.A. Tseytlin and C. Vafa, Nucl. Phys. B372 91992)443;\\
\bibitem{8}I. Antoniadis, C. Bachas, J. Ellis and D.V. Nanopoulos,
Phys. Lett. B211 (1988)393; Nucl. Phys. B328 (1989)117.
\bibitem{9} C. Kounnas and D. L\"ust, Phys. Lett. B 289 (1992)56.
\bibitem{10}  C. Nappi and E. Witten, Phys. Lett. B 293 (1992)309.
\bibitem{11} A. Giveon and A. Pasquinucci, Phys. Lett. B294(1992)162;\\
  A. Giveon and M. Rocek, Nucl. Phys. B 380 (1992)128;\\
 P. Horova,
Phys. Lett. B 278 (1992)101.
\bibitem{KT}  E.W. Kolb and M.S. Turner, {\em ``The Early Universe: Reprints"},
 Addison-Wisley, N.Y., 1988;  {\em ``The Early Universe"}, Addison-Wisley,
N.Y., 1990.
\bibitem{RS} M. Ryan and L. Shepley, {\em ``Homogeneous Relativistic
Cosmologies}, Princeton Univ. Press, Princeton, 1975:\\
D. Kramer {\em et. al.}, {\em ``Exact solutions of Einstein's field
equations"}, Cambridge Univ. Press, Cambridge, 1980.
\bibitem{3'} C.W. Misner, Phys. Rev. Lett. 22 (1969)1071.
\bibitem{D}K. Kikkawa and M. Yamasaki, Phys. Lett. B 149 (1984)357;\\
N. Sakai and I. Senda, Prog. Theo. Phys. 75 (1984)692;\\N. Narain, Phys. Lett.
B 169 (1986)41:\\
T. H. Buscher, Phys. Lett. B 194 (1987)51;\\
A. Giveon, E. Rabinovici and G. Veneziano, Nucl. Phys. B 322
(1989)167.
\bibitem{D'}E. Kiritsis, Mod. Phys. Lett A6 (1991)2871; Nucl. Phys. B 405
(1993)109;\\
M. Rocek and E. Verlinde, Nucl. Phys. B 373 (1992)630; \\
A. Giveon and M. Rocek, Nucl. Phys. B 380 (1992)128;\\
A. Giveon and E. Kiritsis, Nucl. Phys. B 411 (1991)487;\\
A. Giveon, M. Porrati and E. Rabinovici, Phys. Rep. 224 (1994)77;
\\E. Alvarez, L. Alvarez-Gaum\'e and Y. Lozano, {\em ``On non-abelian
duality}, Preprint CERN-TH.7204/94.
\bibitem{Gas} M. Gasperini and R. Ricci, {\em ``Homogeneous conformal
string backgrounds"}, preprint DFTT-02/95, hep-th/9501055.

\end{thebibliography}
\end{document}